%% file: main.tex
\documentclass[acmsmall,screen,nonacm]{acmart}
\usepackage{fancyvrb}
\usepackage{listings}
\usepackage[most]{tcolorbox}
\usepackage{xspace}
\usepackage{enumitem}
\usepackage{siunitx}

\definecolor{leankeyword}{RGB}{0,82,204}
\definecolor{leantype}{RGB}{38,127,153}
\definecolor{leancomment}{RGB}{0,128,0}
\definecolor{leannumber}{RGB}{105,105,105}

\lstdefinelanguage{Lean}{
  sensitive=true,
  morecomment=[l]{--},
  morestring=[b]",
  morekeywords={def,theorem,lemma,where,if,then,else,match,with,by,have,
                let,in,fun,forall,exists,inductive,structure,namespace,end,
                open,import},
  morekeywords=[2]{Type,Prop,Nat,Int,List,LinearOrder,Iff,Membership,And}
}

\lstdefinestyle{leanfigure}{
  language=Lean,
  basicstyle=\ttfamily\fontsize{6.2}{7.2}\selectfont,
  keywordstyle=\color{leankeyword}\bfseries,
  keywordstyle=[2]\color{leantype},
  commentstyle=\color{leancomment},
  stringstyle=\color{red!65!black},
  numbers=none,
  frame=single,
  framerule=0.35pt,
  rulecolor=\color{black!65},
  framesep=3pt,
  xleftmargin=0pt,
  framexleftmargin=0pt,
  framexbottommargin=0pt,
  aboveskip=0pt,
  belowskip=0pt,
  columns=fullflexible,
  keepspaces=true,
  showstringspaces=false,
  breaklines=true,
  tabsize=2
}

\newcounter{CodePairIndex}
\newcounter{CodePairListing}
\newsavebox{\CodePairColumn}
\newlength{\CodePairExtra}
\newlength{\CodePairApplied}
\newcommand{\CodePairRecord}[3]{%
  \csgdef{codepair@height@#1}{#2}%
  \csgdef{codepair@last@#1}{#3}}
\newcommand{\CodePairColumnBegin}[2]{%
  \edef\CodePairKey{\arabic{CodePairIndex}#1}%
  \setcounter{CodePairListing}{0}%
  \setlength{\CodePairExtra}{0pt}%
  \global\CodePairApplied=0pt\relax
  \ifcsdef{codepair@height@\CodePairKey}{%
    \ifcsdef{codepair@height@\arabic{CodePairIndex}#2}{%
      \setlength{\CodePairExtra}{\dimexpr
        \csuse{codepair@height@\arabic{CodePairIndex}#2}-
        \csuse{codepair@height@\CodePairKey}\relax}%
      \ifdim\CodePairExtra<0pt \setlength{\CodePairExtra}{0pt}\fi
    }{}%
  }{}%
  \begin{lrbox}{\CodePairColumn}\begin{minipage}[t]{0.484\linewidth}}
\makeatletter
\newcommand{\CodePairColumnEnd}{%
  \end{minipage}\end{lrbox}%
  \immediate\write\@auxout{\string\CodePairRecord{\CodePairKey}%
    {\the\dimexpr\ht\CodePairColumn+\dp\CodePairColumn-\CodePairApplied\relax}%
    {\arabic{CodePairListing}}}%
  \usebox{\CodePairColumn}}
\makeatother
\lstnewenvironment{LeanCode}[1][]
  {\lstset{style=leanfigure,#1}%
   \ifdefined\CodePairKey
     \stepcounter{CodePairListing}%
     \ifcsdef{codepair@last@\CodePairKey}{%
       \ifnum\value{CodePairListing}=\csuse{codepair@last@\CodePairKey}\relax
         \lstset{framexbottommargin=\CodePairExtra}%
         \global\CodePairApplied=\CodePairExtra
       \fi
     }{}%
   \fi}
  {}

\newenvironment{CodePair}
  {\stepcounter{CodePairIndex}\par\noindent\ignorespaces}
  {\par}
\newenvironment{CodePairLeft}
  {\CodePairColumnBegin{L}{R}}
  {\CodePairColumnEnd\hspace{0.03\linewidth}\ignorespacesafterend}
\newenvironment{CodePairRight}
  {\CodePairColumnBegin{R}{L}}
  {\CodePairColumnEnd\ignorespacesafterend}

\newcommand{\TableStyle}{%
  \small
  \setlength{\tabcolsep}{4pt}%
  \renewcommand{\arraystretch}{1.1}%
}
\newcommand{\goodchange}[1]{\textcolor{green!50!black}{\scriptsize (#1)}}
\newcommand{\badchange}[1]{\textcolor{red!70!black}{\scriptsize (#1)}}
\newcommand{\name}{\textsc{VeriBase}\xspace}

\newtcolorbox{rqsummary}[1][Summary]{
  enhanced,
  title={#1},
  colback=black!3,
  colframe=black!58,
  colbacktitle=black!58,
  coltitle=white,
  fonttitle=\small\bfseries,
  boxrule=1.2pt,
  arc=5pt,
  outer arc=5pt,
  left=7pt,
  right=7pt,
  top=6pt,
  bottom=6pt,
  toptitle=0.0pt,
  bottomtitle=0.0pt,
  before skip=4pt,
  after skip=4pt
}
\AtBeginDocument{%
  }

\setcopyright{none}
\begin{document}

\title{\name: Automated Construction of Theorem Libraries for Software Verification
}

\input{authors}

\begin{abstract}

\input{tex/000_abs}

\end{abstract}

\ccsdesc[500]{Software and its engineering~Formal software verification}

\keywords{formal verification, theorem libraries, large language models, Lean}

\maketitle

\input{tex/001_intro}

\input{tex/002_background}

\input{tex/003_method}

\input{tex/004_library}
\input{tex/005_experiments}

\input{tex/006_threats}
\input{tex/007_related_work}
\input{tex/008_conclusion}

\section{Data Availability}
The code and generated libraries are publicly available at
\url{https://github.com/zzhisthebest/Veribase}.

\bibliographystyle{ACM-Reference-Format}
\bibliography{tex/references}

\end{document}

%% file: authors.tex
\author{Zihao Zhang}
\email{zihaozhang@smail.nju.edu.cn}
\affiliation{%
  \institution{Nanjing University}
  \city{Nanjing}
  \country{China}
}

\author{Zenan Li}
\email{lizn@smail.nju.edu.cn}
\affiliation{%
  \institution{Nanjing University}
  \city{Nanjing}
  \country{China}
}

\author{Taolue Chen}
\email{t.chen@bbk.ac.uk}
\affiliation{%
  \institution{Birkbeck, University of London}
  \city{London}
  \country{United Kingdom}
}

\author{Zhe Zhou}
\email{zhouzhe956@pku.edu.cn}
\affiliation{%
  \institution{Peking University}
  \city{Beijing}
  \country{China}
}

\author{Tao Xie}
\email{taoxie@pku.edu.cn}
\affiliation{%
  \institution{Peking University}
  \city{Beijing}
  \country{China}
}
\affiliation{%
  \institution{Fudan University}
  \city{Shanghai}
  \country{China}
}

\author{Yuan Yao}
\email{y.yao@nju.edu.cn}
\affiliation{%
  \institution{Nanjing University}
  \city{Nanjing}
  \country{China}
}

\renewcommand{\shortauthors}{Zhang et al.}

%% file: tex/000_abs.tex
Software verification is typically more challenging than coding. 
Software developers routinely build on mature libraries and APIs, but turning successful verification solutions into reusable libraries still requires substantial expert effort.
In this paper, we propose \name, a framework for automatically constructing software-verification libraries by jointly generalizing code and its associated theorems from successful verification solutions.
The key challenge is to generalize implementations together with the theorem statements and proofs that depend on these implementations, and reconcile overlapping components across verification tasks.
To this end, \name treats code and its associated theorems as declaration families, and designs a three-phase pipeline, which proves seed tasks, generalizes code and associated theorems, and harvests the resulting families into shared interfaces. \name  updates dependent declarations as interface change and uses Lean to check the resulting code and proofs.
We evaluate \name across four benchmarks with three backend LLMs. The results show that the general-purpose library built from independent 4,191 programs increases the average task success rate by up to 17.0\% and meanwhile reduces the proof completion time by up to 21.4\%. When \name constructs a benchmark-specific library, further improvement can be observed. %

%% file: tex/001_intro.tex
\section{Introduction}

As coding tools and development practices based on a large language model (LLM)---such as vibe coding---gain widespread adoption, %
more software is being generated or modified by LLMs. While this trend can substantially increase software development productivity, it also amplifies a longstanding challenge of assuring that increasingly large volumes of 
software behave correctly. LLM-generated code may appear plausible and pass available tests while still containing subtle bugs. Testing remains essential, but it alone generally cannot establish correctness across all possible inputs and execution paths.

Formal verification offers a complementary approach to software quality assurance, providing rigorous guarantees about program behavior by expressing program specifications in formal languages and using machine-checked proofs to establish that programs satisfy the specifications. Interactive theorem provers (ITPs) such as Lean, Rocq (formerly Coq), and Isabelle provide environments for constructing and checking proofs of program correctness. Long before the recent rise of LLM-based code generation, ITPs had been successfully applied to high-assurance software, including seL4~\cite{klein2009sel4}, an operating-system kernel formally verified in Isabelle, and CompCert~\cite{leroy2016compcert}, a C compiler formally verified in Rocq.

However, proof development is often substantially more expensive than software development for two main reasons: the complexity of verification tasks and the limited infrastructure for proof reuse.
First, \emph{verification tasks are intrinsically demanding}: implementing a program requires specifying how to perform a computation, whereas verifying the program additionally requires establishing that the implementation satisfies its specification across all relevant executions.
Consequently, the proof artifact can be considerably larger than the code that the proof artifact verifies. For instance, the seL4 microkernel verification required over 200,000 lines of Isabelle proof scripts to verify about 7,500 lines of C code~\cite{klein2009sel4}.
Recent advances in LLM-based formal reasoning, including neuro-symbolic proof search~\cite{he2026neurosymbolic}, proof planning~\cite{zhang2026planninghammerdifficultyawaredecomposition}, and hierarchical proof search~\cite{li2026goedelcodeproverhierarchicalproofsearch}, among others~\cite{yang2023leandojo,thakur2023context,xin2024deepseek}, have improved the automation of proof development, making software verification increasingly practical. Nevertheless, generating proofs remains substantially more costly than generating code.

Second, \emph{proof reuse has limited infrastructure}, with library construction and maintenance relying heavily on expert effort. Modern code development benefits from a mature ecosystem of reusable libraries and APIs, such as the C++ Standard Library and Python's standard library, reducing the need to implement common functionality from scratch. In contrast, theorem libraries for software verification remain comparatively scarce~\cite{lammich2010isabelle,barrett2026cslib}. Successful verification efforts do not automatically translate into reusable libraries: experts must identify reusable functionality, design general code interfaces and associated theorems, and integrate these declarations with existing libraries. This work requires expertise beyond completing an individual verification task and must be repeated as new library functionality is needed.

\begin{figure}[t]
    \centering
    \includegraphics[width=0.95\linewidth]{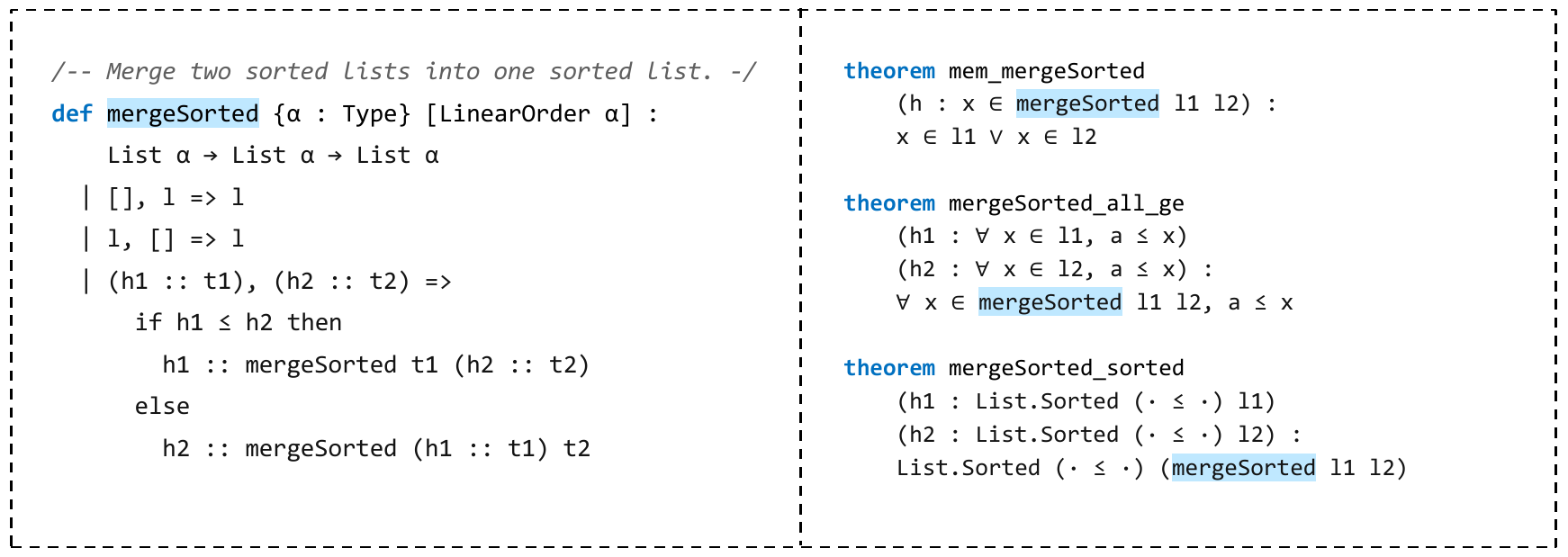}
    \vspace{-2mm}
    \caption{An example of a code snippet and its surrounding theorems. The
    three theorems (i.e., \texttt{mem\_mergeSorted},
    \texttt{mergeSorted\_all\_ge}, and \texttt{mergeSorted\_sorted}) are all
    stated over \texttt{mergeSorted}.}
    \label{fig:code-snippets}
    \vspace{-2mm}
\end{figure}

These limitations motivate a question: can we \textit{automatically} turn successful software verification solutions into reusable theorem libraries, reducing the manual effort of library construction and allowing future tasks to benefit from previous verification work?
Automating this process presents two main challenges.
First, \emph{code and theorems must be generalized together}. Successful
solutions often contain task-specific implementations; generalizing these
implementations requires adapting dependent theorem statements and proofs.
For example, the theorems in
Fig.~\ref{fig:code-snippets} refer to \texttt{mergeSorted}; changing its interface
requires updating the theorems and their uses.
Second, \emph{overlapping components must be integrated across tasks}.
Declarations from different solutions must be reconciled into shared interfaces
with consistent dependencies, rather than merely collected into a library.

\smallskip
\noindent\textbf{Our work.} To address these challenges, we propose \name, a framework for automatically constructing theorem libraries from verification tasks expressed in Lean.
It treats code and its associated theorems as a \emph{declaration family}. Its
\emph{code-first, theorems-second} principle makes interface changes explicit:
each operation first handles the code and then adapts the dependent theorem
statements and proofs. The theorem prover (e.g., Lean) checks the resulting code and proofs.
\name constructs libraries in three phases. \emph{Proving} produces successful
solutions through multiple rounds of task solving. \emph{Generalization} extracts
computational components and intermediate proof facts, generalizes their
interfaces, and updates dependent declarations. \emph{Harvesting} integrates
candidates across tasks by adding new functionality and properties, skipping
covered candidates, and merging overlapping interfaces while updating affected
declarations and imports.

We evaluate the effectiveness of the libraries constructed by \name in
improving task success rates and reducing completion time in Lean~\cite{moura2021lean}, with four benchmarks and three backend LLMs. On
Verina~\cite{ye2025verina},
FVAPPS~\cite{dougherty2025proving}, and
VeriContest~\cite{xie2026vericontest}, the general library, constructed from
tasks outside these benchmarks, improves the average task success rate from
63.0\% to 73.0\% and reduces average completion time by 13.2\%. These results demonstrate that the constructed library
serves as a general-purpose library for software verification across benchmarks.
When additional tasks from the target benchmark's domain are available, \name
can construct a benchmark-specific library from these tasks. %
Combining the general and benchmark-specific libraries further increases the
average task success rate to 78.5\% and reduces average completion time by
27.9\% relative to the no-library baseline.
On the repository-scale benchmark VeriSoftBench~\cite{xin2026verisoftbench}, after removing human-developed
supporting theorems (which serve a role analogous to our constructed library), our constructed library improves the average task success
rate across three backend LLMs from 56.3\% to 77.0\%. This performance is close to that achieved with the original supporting theorems (e.g., with GLM 5.2, \name library achieves 81.0\% success rate versus 84.0\% with the original supporting theorems), suggesting comparable quality
between automatic and human-developed libraries for repository-level verification.

In summary, this paper makes the following main  contributions.
\begin{itemize}[leftmargin=*]
\item \textbf{Automated construction of code and theorem libraries.}
We propose \name, a framework for  jointly generalizing implementations and their associated
theorems from successful verification solutions, updating dependent declarations,
and consolidating overlapping families across tasks into reusable interfaces.

\item \textbf{New Lean~4 libraries for software verification.}
We construct a general library containing 1,609 code declarations and 3,239
theorems, together with benchmark-specific libraries that provide additional
code and theorems for their task distributions and repositories.

\item \textbf{An empirical evaluation of downstream reuse.} We evaluate
the libraries across four benchmarks and three backend LLMs. Comparisons on
VeriContest with raw-solution retrieval and ablated variants examine the benefits
of retaining code with its associated theorems, generalization, and library integration.

\end{itemize}

%% file: tex/002_background.tex
\section{Background}
\subsection{Software Verification}

Software verification aims to check whether software satisfies specified requirements or properties typically described as a specification. 
For example, a sorting function must return an ordered permutation of its input.
Verification tools vary in how proof obligations are discharged and in the degree of automation that these tools provide.
Tools such as Dafny~\cite{leino2010dafny}, Verus~\cite{lattuada2023verus}, Frama-C/WP~\cite{baudin2020wp}, and Viper~\cite{muller2015viper} emphasize automated deductive verification: they generate verification conditions from programs and specifications, and rely heavily on automated solvers, particularly SMT solvers, to discharge verification conditions. This approach can provide a high degree of automation when the resulting verification conditions fall within the capabilities of the underlying solvers. 

Interactive theorem provers (ITPs), such as Lean~\cite{moura2021lean}, Rocq~\cite{bertot2013interactive}, and Isabelle~\cite{nipkow2002isabelle}, instead provide expressive logical frameworks in which users can construct and check proofs, while also incorporating substantial proof automation. This flexibility supports specifications and proofs that require reasoning beyond what can be discharged automatically, at the cost of greater proof-development effort.
In Lean, a software verification task typically includes code to be implemented,
a specification describing its intended behavior, and a target theorem stating
that the code satisfies the specification. Completing the task requires both
an implementation and a correctness proof. Constructing this proof often
involves establishing intermediate properties of individual computational
steps or helper functions, and using these properties to prove the target
theorem. Lean's kernel type-checks the resulting proof term. A successful solution may therefore contain not only code and its final correctness proof, but also intermediate theorems about the code.

\subsection{Theorem Libraries for Software Verification}
Theorem libraries organize reusable definitions and theorems to support new proofs.
In software verification, theorems typically state properties of code, naturally giving rise to a code-centered structure where 
code and its associated theorems form declaration families. 
When code is reused in a new verification task, %
these associated theorems provide reusable reasoning about its behavior without requiring the same properties to be established from scratch. %
Representative verification libraries include Rocq's \texttt{std++}~\cite{stdpp}, the Isabelle Collections Framework~\cite{lammich2010isabelle}, Verus's \texttt{vstd}~\cite{lattuada2023verus},
Dafny's standard libraries~\cite{dafnystandardlibraries},  and
Lean's CSLib~\cite{barrett2026cslib}. %

Software verification can significantly benefit from theorem libraries, especially in the context of LLM-based software verification. Libraries provide code that LLMs can
incorporate into implementations and theorems that they can invoke in
correctness proofs. This reuse reduces the need to generate code from scratch or to reestablish properties already captured by the library, allowing  
LLMs to focus on adapting reusable components and proving the remaining task-specific
properties. 
In addition, existing libraries can also serve as valuable corpora for training theorem-proving models, exposing them to formal syntax, proof patterns, and declaration usage. For example, Lean~4's
\texttt{Mathlib}~\cite{mathlib2020} has been used as training data for neural theorem-proving systems%
~\cite{yang2023leandojo,xin2024deepseek}. 

Despite useful, turning successful task-level solutions into library components typically requires substantial manual effort. 
For given verification projects that contain useful code and
theorems, one must separate task-specific details, generalize interfaces, and organize dependencies. Completed verification tasks provide
valuable material for library construction, but knowledge reuse
requires transforming task-specific code and theorems into coherent library
components.

%% file: tex/003_method.tex
\section{Methodology of \name}

\name is guided by the observation that reusable knowledge in software
verification is rarely an isolated theorem. Code provides the computational
semantics, while its associated theorems provide an interface for reasoning
about the code's  behavior %
without repeatedly unfolding the implementation. \name therefore
treats code and its associated theorems as a \emph{code-centered declaration
family}, rather than as independent library items. Since code forms the backbone
of each family, \name follows a \emph{code-first, theorems-second} principle:
each operation first handles the code and then its associated theorems.

Transforming completed tasks into such families requires abstraction at two
levels: within each task and across tasks. Within each task, \name generalizes a successful proof by removing task-specific structure while
preserving the essential specification and verification goal. 
Across tasks,
\name compares candidate families against the growing library,
consolidates overlapping interfaces, and organizes the retained knowledge into
coherent modules.

\begin{figure}[t]
    \centering
    \includegraphics[width=0.90\linewidth]{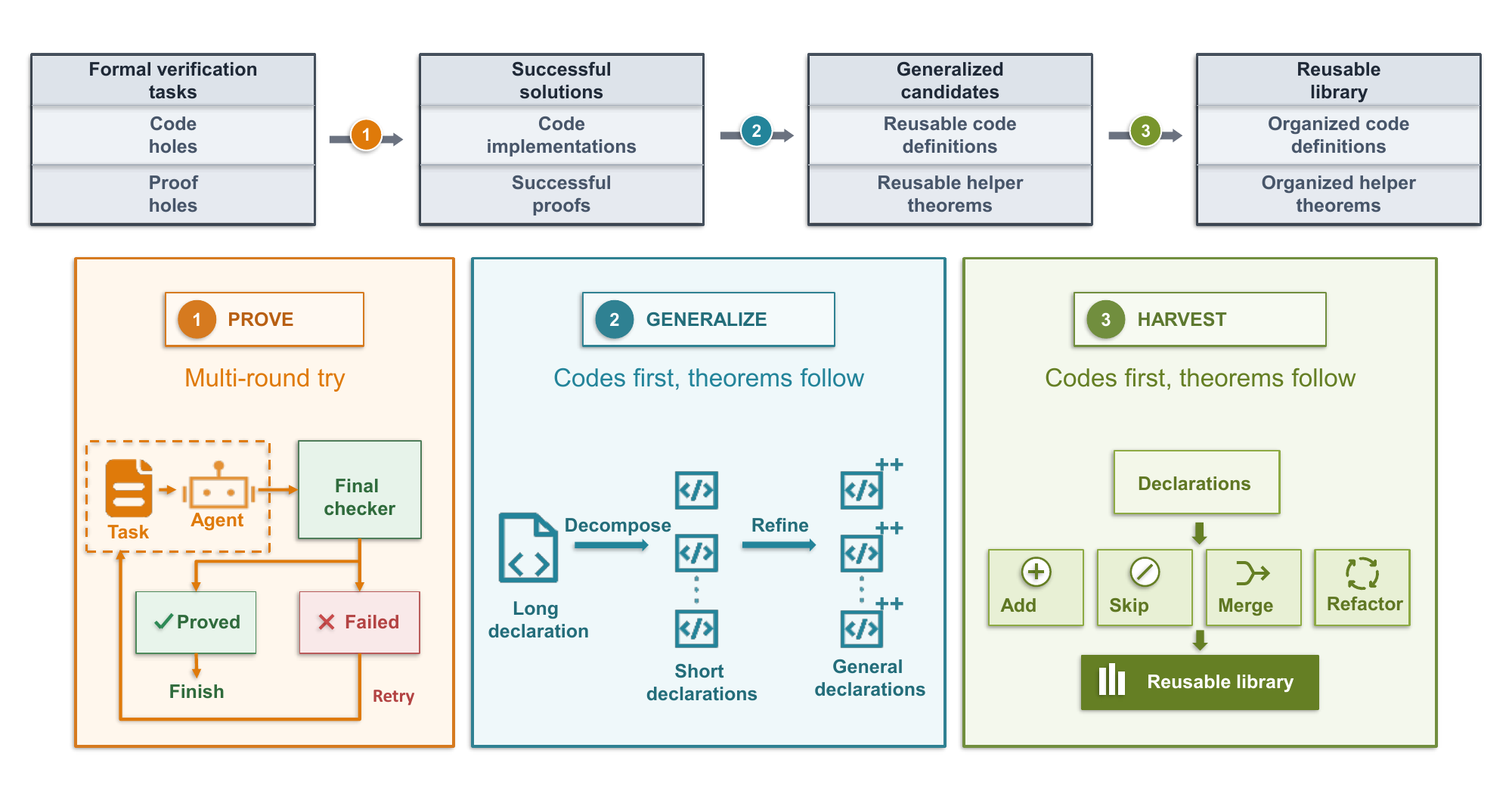}
    \vspace{-2mm}
    \caption{Overview of \name. \name proves software-verification tasks to
    obtain Lean-validated solutions, generalizes them into code-centered
    declaration families, and integrates these families into a coherent library
    through addition, skipping, merging, and library-level refactoring.}
    \Description{A workflow diagram showing \name completing tasks selected for library construction, generalizing successful solutions, and harvesting reusable declarations into a theorem library.}
    \label{fig:veribase-framework}
    \vspace{-2mm}
\end{figure}

An overview of \name is shown in Figure~\ref{fig:veribase-framework}, which demonstrates how \name realizes this
two-level abstraction in three phases: \emph{proving} produces Lean-validated seed solutions;
\emph{generalization} turns each successful solution into reusable code and associated theorems;
and \emph{harvesting} integrates the resulting declaration families into a coherent
library.

\subsection{Proving Phase}
The proving phase produces the raw material. %
For each code-verification task, a formal specification written in Lean is given as input, and the goal is two-fold:  
implementing the code according to the formal specification and
proving the theorem that the resulting code satisfies the formal specification. 
\name uses a multi-round strategy to solve as many tasks as possible: failed tasks are
retried in subsequent rounds until they succeed or the budget is exhausted.
Successful tasks are collected and passed to the next phase.

For each task in a round, \name follows a code-first, proof-second workflow.
The agent used by \name first implements the code according to the formal specification and
invokes Lean to check the implementation. %
If the check fails, the agent revises the code based
on Lean's diagnostics, forming a generate-check-repair loop that continues
until the implementation is accepted by Lean. 
The agent then proves the target theorem using
the same iterative generate-check-repair process until the entire task is verified
by Lean. If the proof process becomes stuck, \name prompts the agent to
revisit and modify the implementation and reconstruct the proof accordingly. 
After the agent finishes processing a task in a
round, \name performs a final verification of the complete task file using Lean. 
A task is considered successful only if the file compiles
successfully and contains no unproved placeholders or unauthorized assumptions.
A task that times out or
fails validation is considered failed. Section~\ref{sec:experimental-setup}
details the validation criteria used to determine task success. %

\subsection{Generalization Phase}
The generalization phase produces reusable code and theorems as candidates
for harvesting through two stages: \emph{code generalization} and
\emph{theorem generalization}.

\begin{figure}[t]
\centering
\textbf{Code refactoring}\par\smallskip
\begin{CodePair}
\begin{CodePairLeft}
\begin{LeanCode}
def lastDigit (n : Int) : Nat :=
  (Int.natAbs n) %

def sort_by_last_digit (numbers : List Int)
  (h_precond : sort_by_last_digit_precond numbers) :
  List Int :=
  sort numbers
where
  insert (x : Int) : List Int -> List Int
    | [] => [x]
    | y :: ys =>
      if lastDigit x <= lastDigit y
      then x :: y :: ys
      else y :: insert x ys
  sort : List Int -> List Int
    | [] => []
    | x :: xs => insert x (sort xs)
\end{LeanCode}
\end{CodePairLeft}
\begin{CodePairRight}
\begin{LeanCode}
def lastDigit (n : Int) : Nat :=
  (Int.natAbs n) %

def insertByLastDigit
  (x : Int) : List Int -> List Int
  | [] => [x]
  | y :: ys =>
    if lastDigit x <= lastDigit y
    then x :: y :: ys
    else y :: insertByLastDigit x ys

def sortByLastDigit : List Int -> List Int
  | [] => []
  | x :: xs =>
    insertByLastDigit x (sortByLastDigit xs)

def sort_by_last_digit (numbers : List Int)
  (h_precond : sort_by_last_digit_precond numbers) :
  List Int :=
  sortByLastDigit numbers
\end{LeanCode}
\end{CodePairRight}
\end{CodePair}

\medskip
\textbf{Code refinement}\par\smallskip
\begin{CodePair}
\begin{CodePairLeft}
\begin{LeanCode}
def insertByLastDigit
  (x : Int) : List Int -> List Int
  | [] => [x]
  | y :: ys =>
    if lastDigit x <= lastDigit y
    then x :: y :: ys
    else y :: insertByLastDigit x ys
\end{LeanCode}
\end{CodePairLeft}
\begin{CodePairRight}
\begin{LeanCode}
def insertByKey {A B : Type} [LinearOrder B]
  (key : A -> B) (x : A) : List A -> List A
  | [] => [x]
  | y :: ys =>
    if key x <= key y
    then x :: y :: ys
    else y :: insertByKey key x ys
\end{LeanCode}
\end{CodePairRight}
\end{CodePair}
\vspace{-1mm}
\caption{An example of code generalization (\S\ref{sect:cg}). Code refactoring extracts
the insertion and sorting functions as separate declarations.
Code refinement replaces the fixed last-digit sorting criterion
with a key-function parameter.}
\Description{Two before-and-after Lean code examples showing code refactoring and code refinement.}
\label{fig:code-generalization-examples}
\vspace{-2mm}
\end{figure}

\subsubsection{Code Generalization} \label{sect:cg}
Code generalization transforms task-specific implementations into reusable
code. This step is necessary: if the code remains tied to one specific 
task, the theorems stated over the code remain difficult to reuse. 
\name applies two
mechanisms for code generalization: code refactoring and code refinement.
Throughout code generalization, \name keeps the original code specification
unchanged and re-verifies that the generalized implementation satisfies the specification.
Doing so ensures that the generalized code continues to %
satisfy the requirements of the  original specification.

Code refactoring splits a complex task implementation into smaller, potentially
reusable computational components while preserving the properties required by
the original specification.
\name instructs the  agent to read the complete successful solution and identify
computational steps (within large functions) that can be called independently
and composed to implement the original functionality. For each selected step,
the agent turns its external data dependencies into explicit parameters,
extracts it as a named \texttt{def}, and replaces the original code fragment
with a call to that function.
After refactoring, \name invokes Lean to verify the refactored implementation still satisfies the original specification; only
refactored code that successfully passes this verification proceeds to subsequent processing.

Code refinement further abstracts each refactored function into a more reusable form. The
core principle is to abstract away task-specific details while preserving the
original semantics, thereby making the function applicable to a broader class of problems. 
Specifically, \name instructs the agent to
inspect the refactored function and identify task-specific details in its interface
and implementation, such as fixed types or operations. The agent turns these details into explicit parameters and generalizes the function signature and body accordingly. 
A refinement is accepted only if the refined code subsumes the original
functionality: instantiating its parameters with the original types and operations must recover the original behavior.
The agent then updates dependent declarations and proofs to use the generalized
interface and invokes Lean to validate the refined implementation and updated proofs.

\begin{example}
Figure~\ref{fig:code-generalization-examples} illustrates how refactoring
extracts insertion and sorting functions from \texttt{sort\_by\_last\_digit}.
Refinement then generalizes the insertion function to arbitrary element types
and key functions. Instantiating these parameters with integers and the
last-digit operation recovers the original behavior.
\end{example}

\begin{figure}[t]
\centering
\textbf{Proof refactoring}\par\smallskip
\begin{CodePair}
\begin{CodePairLeft}
\begin{LeanCode}
theorem sort_by_last_digit_postcond_satisfied
  (numbers : List Int) (h_precond : precond numbers) :
  postcond numbers (sort_by_last_digit numbers h_precond)
    h_precond := by
  ...
  have h_perm : (sortByLastDigit numbers).Perm numbers := by
    ...
  have h_pairwise : List.Pairwise (fun a b =>
    lastDigit a <= lastDigit b) (sortByLastDigit numbers) := by
    ...
  ...
\end{LeanCode}
\end{CodePairLeft}
\begin{CodePairRight}
\begin{LeanCode}
theorem sortByLastDigit_perm ...

theorem sortByLastDigit_pairwise ...

theorem sort_by_last_digit_postcond_satisfied
  (numbers : List Int) (h_precond : precond numbers) :
  postcond numbers (sort_by_last_digit numbers h_precond)
    h_precond := by
  ...
  have h_perm : (sortByLastDigit numbers).Perm numbers :=
    sortByLastDigit_perm numbers
  have h_pairwise : List.Pairwise (fun a b =>
    lastDigit a <= lastDigit b) (sortByLastDigit numbers) :=
    sortByLastDigit_pairwise numbers
  ...
\end{LeanCode}
\end{CodePairRight}
\end{CodePair}

\medskip
\textbf{Theorem refinement}\par\smallskip
\begin{CodePair}
\begin{CodePairLeft}
\begin{LeanCode}
theorem insertByLastDigit_perm
  (x : Int) (ys : List Int) :
  (insertByLastDigit x ys).Perm (x :: ys)
\end{LeanCode}
\end{CodePairLeft}
\begin{CodePairRight}
\begin{LeanCode}
theorem insertByKey_perm
  {A B : Type} [LinearOrder B]
  (key : A -> B) (x : A) (ys : List A) :
  (insertByKey key x ys).Perm (x :: ys)
\end{LeanCode}
\end{CodePairRight}
\end{CodePair}
\vspace{-1mm}
\caption{An example of theorem generalization (\S\ref{sect:tg}). Proof refactoring extracts
local permutation and ordering facts into theorems.
Theorem refinement then generalizes the permutation theorem from
last-digit insertion to insertion based on an arbitrary key function.}
\Description{Two before-and-after Lean code examples showing proof refactoring and theorem refinement.}
\label{fig:theorem-generalization-examples}
\vspace{-2mm}
\end{figure}

\subsubsection{Theorem Generalization} \label{sect:tg}
Theorem generalization transforms task-specific proofs into reusable
theorems. This step is necessary because useful proof knowledge is often buried
inside a long proof of a single target theorem. If these intermediate facts
remain local to that proof, other tasks cannot invoke them directly. \name
therefore applies two mechanisms for theorem generalization: proof
refactoring and theorem refinement.
Throughout theorem generalization, \name keeps the original target theorem
statement unchanged.
Doing so ensures that the original verification goal remains satisfied after
generalization.

Proof refactoring extracts intermediate facts from a long task proof into
reusable theorems.
Specifically, \name instructs the agent to inspect local \texttt{have}
statements and other steps in the main proof. It selects intermediate facts
that can be stated using explicit parameters and hypotheses, and applied
without relying on the original task's surrounding proof context, rather than
extracting every proof step. For each selected fact, the agent makes the
variables and premises on which the fact depends explicit in a named theorem and
reconstructs its proof. The agent then replaces the original proof fragment with an
application of that theorem, supplying the corresponding arguments and
hypotheses. Finally, \name invokes Lean to check the refactored proof.

Theorem refinement further rewrites extracted theorems into more
reusable forms. 
Theorem refinement is also performed by the agent. \name
instructs the agent to inspect the hypotheses and conclusion of each auxiliary theorem, and identify task-specific elements that restrict the theorem's applicability. 
Based on
the structure and meaning of the theorem, the agent may remove unnecessary hypotheses or replace %
overly restrictive hypotheses with weaker conditions; lift concrete types, constants, functions, and relations
into parameters; restate properties expressed through task-specific
preconditions or postconditions as direct properties of the code; or adapt the
theorem to a generalized code interface. The agent reconstructs the proofs of the refined auxiliary
theorems and updates any dependent proofs, and then invokes Lean to verify both the
refined theorems and their dependent proofs.

\begin{example} 
As shown in
Figure~\ref{fig:theorem-generalization-examples}, the original correctness
proof for the code in Figure~\ref{fig:code-generalization-examples} establishes permutation and pairwise-ordering properties inline through
\texttt{h\_perm} and \texttt{h\_pairwise}. \name extracts these properties into two auxiliary theorems:
\texttt{sortByLastDigit\_perm} and \texttt{sortByLastDigit\_pairwise}. 
This step of theorem refactoring  is not limited to declarative proof steps (such as
\texttt{have} statements); reusable properties within procedural proof fragments can also be extracted into
theorems.
Theorem refinement then generalizes the extracted theorems. Namely,   
\texttt{insertByLastDigit\_perm} is refined into
\texttt{insertByKey\_perm}, replacing the task-specific insertion function with
the generalized key-based insertion function.
\end{example}

\subsubsection{Connecting Code and Theorems}
Following the principle that code forms the backbone and theorems are organized
around it, \name first generalizes the code and then generalizes the
theorems, preserving their dependency
relationships. 
The association between code and theorem is determined by references to code declarations in
theorem statements: a function may have multiple associated theorems, and a
theorem may describe properties of several functions composed together.
Standalone theorems reference no code in the VeriBase library;
instead, they rely on only declarations from Mathlib and Lean's
standard libraries.

\subsection{Harvesting Phase} \label{sect:hav}
The harvesting phase incrementally integrates generalized candidates into a
coherent Lean library. It starts by extracting and filtering candidate declarations. For each task file produced by the generalization phase, \name invokes the custom Lean executable
\texttt{extract\_decls} through \texttt{lake exe} to extract code and theorem declarations, while
excluding declarations that encode the task specification. For each extracted
declaration, \name generates a short  description and embeds it using
\texttt{all-MiniLM-L6-v2}~\cite{allminilml6v2}.
We define the cross-task frequency of a candidate as the number of
distinct task files containing a declaration whose description embedding has similarity above a predefined threshold to the
candidate. Candidates below a minimum frequency
are removed before library integration.

The remaining candidates are then integrated into %
the library in source-task batches. Grouping candidates by their source tasks preserves the context needed to integrate mutually
dependent code and theorems together. For each batch, the harvester reads the
complete source declarations, identifies their dependency groups, and processes
code before the theorems that depend on the code. 
The harvester  then inspects the current library by traversing its module hierarchy and searching declaration names and contents to
locate potentially overlapping interfaces. Batches are processed sequentially,
so each batch is compared against the library resulting from all previously processed batches.

When integrating a candidate group, the harvester applies three types of actions to its declarations, i.e., Add, Skip, and Merge.

\begin{itemize}[leftmargin=*]
    \item {\em Add} is used when the candidate provides functionality or a
    property not already represented in the library. The harvester may generalize the candidate's 
    interface with respect to existing declarations, rename the candidate, and place it
    in the appropriate namespace and module together with its dependencies.
    
    \item {\em Skip} is used when an existing declaration is equivalent to or
    more general than the candidate. A candidate is also skipped if it remains tied to its source task and cannot be generalized without changing its behavior, or if it captures only a trivial property.  
    
    \item {\em Merge} is used when existing declarations overlap with the
    candidate but do not subsume it. The harvester compares code
    by its computational behavior and theorems by their statements,and then
    incorporates the functionality or properties into the existing
    interface and updates dependent declarations.
\end{itemize}

Figure~\ref{fig:harvesting-merge-example} shows a typical example. A task
produces a one-sided inclusive-range definition, together with theorems about
membership and duplicate freeness. Since the library already contains a more
general two-sided interval abstraction, the harvester does not add a parallel
one-sided API. Instead, the harvester skips the narrower definition and generalizes the reusable 
theorems to the two-sided intervals before merging them into the existing declaration family.

\begin{figure}[t]
\centering
\begin{CodePair}
\begin{CodePairLeft}
\centering
\textbf{Library before merging}\par\smallskip
\begin{LeanCode}
-- Existing abstraction in the library
def inclusiveRange (l r : Nat) : List Nat :=
  (List.range (r + 1 - l)).map
  (fun d => l + d)
\end{LeanCode}
\textbf{Candidates}\par\smallskip
\begin{LeanCode}
-- Candidate family from one task
def inclusiveRange (n : Nat) : List Nat :=
  List.range (n + 1)

theorem mem_inclusiveRange_iff
  (x n : Nat) :
  Iff (Membership.mem x (inclusiveRange n))
    (x <= n) := by
  ...

theorem nodup_inclusiveRange
  (n : Nat) :
  (inclusiveRange n).Nodup := by
  ...
\end{LeanCode}

\end{CodePairLeft}
\begin{CodePairRight}
\centering
\textbf{Library after merging}\par\smallskip
\begin{LeanCode}
-- Existing abstraction is kept
def inclusiveRange (l r : Nat) : List Nat :=
  (List.range (r + 1 - l)).map
  (fun d => l + d)

@[simp] theorem mem_inclusiveRange_iff
  {l r x : Nat} :
  Iff (Membership.mem x (inclusiveRange l r))
    (And (l <= x) (x <= r)) := by
  ...

@[simp] theorem nodup_inclusiveRange
  {l r : Nat} :
  (inclusiveRange l r).Nodup := by
  ...
\end{LeanCode}
\end{CodePairRight}
\end{CodePair}
\vspace{-1mm}
\caption{An example of merging during harvesting. Since the library
already contains code for a two-sided range, the harvester skips the narrower
one-sided range code and generalizes its surrounding theorems
to the existing two-sided range code.}
\Description{}
\label{fig:harvesting-merge-example}
\vspace{-2mm}
\end{figure}

Adding or merging a candidate may also trigger library-level refactoring. For
example, the harvester may extract code shared by several declarations, rewrite
existing declarations against a more general interface, or move related
declarations into a common module. The
harvester updates affected dependencies and imports
whenever an interface or module changes. After processing a batch, the
harvester compiles the affected modules; unsuccessful edits are repaired or
rolled back before processing continues. Finally, \name compiles the complete
library and checks that it contains no unproved placeholders or unauthorized
axioms. Thus, only successfully integrated batches contribute to the resulting
library.

%% file: tex/004_library.tex
\section{Generated Libraries}
\label{sec:veribase-library}

The general library is constructed from 4,191 code-verification tasks
from LeetCode and OpenCodeInstruct~\cite{ahmad2025opencodeinstruct},
translated into Lean~4. The proving phase verifies 3,835 tasks.
Generalization and frequency filtering produce 6,064 code implementations and 6,359 theorem
candidates; harvesting yields 1,609 code declarations and 3,239 theorems,
totaling 4,848 declarations. Counts include only explicit source declarations,
excluding compiler-generated ones.

Harvesting uses a similarity threshold of 0.8 and a minimum frequency of 3.
We use Claude Code as the agent harness, which manages tool use and interaction with the verification environment, and
DeepSeek V4 Pro~\cite{xu2026deepseek} as the backend LLM. All proving,
generalization, and harvesting decisions are automated. \name is
implemented in Python and Lean~4.

The general library supports reuse across tasks; benchmark-specific libraries
use the same pipeline with subsets of each benchmark, as detailed in
Section~\ref{sec:experimental-setup}, and are tailored to the benchmark's  task distribution.
Each library is compiled as a single Lean project and automatically checked
for unresolved proof placeholders (\texttt{sorry} or \texttt{admit}) and new axioms.

\subsection{General Library}

\begin{figure}[t]

    \centering
    \includegraphics[width=0.415\linewidth]{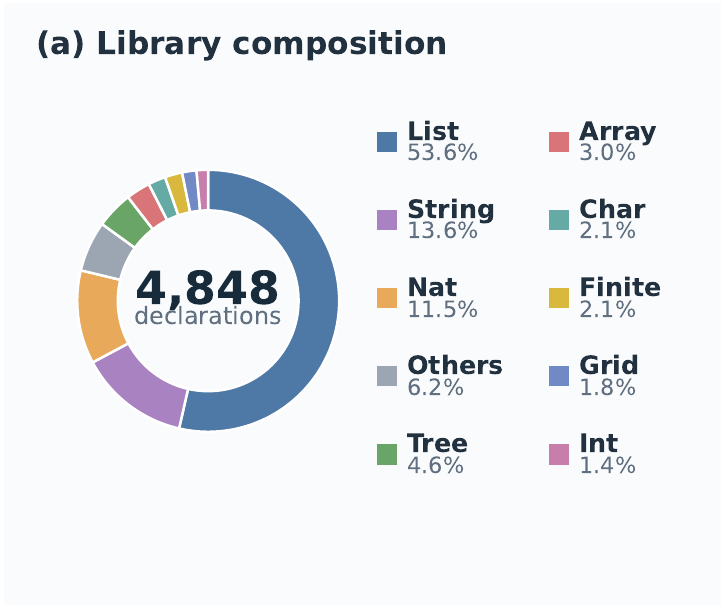}
    \hfill
    \includegraphics[width=0.565\linewidth]{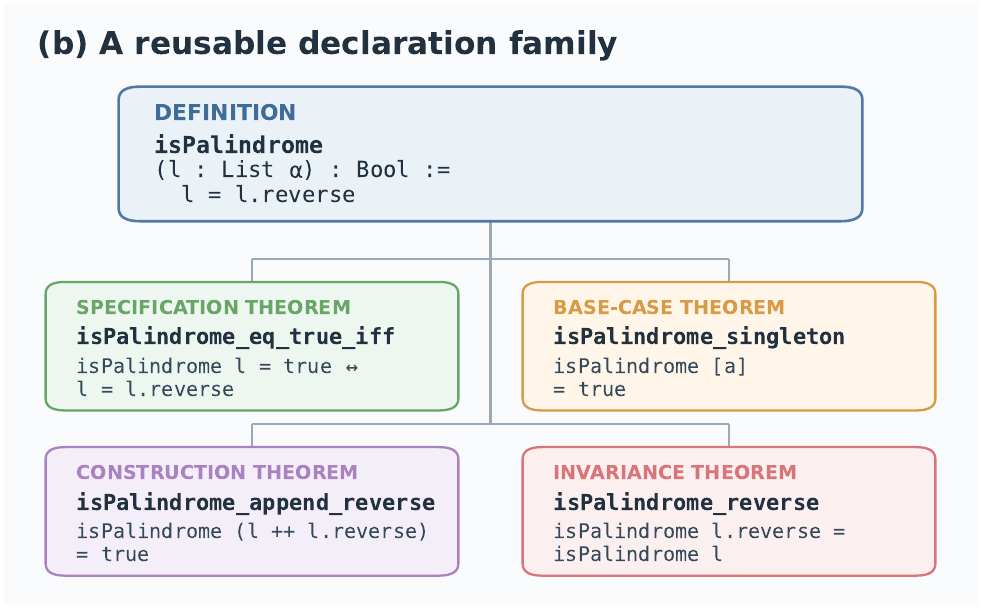}
    \vspace{-2mm}
    \caption{Organization and code-centered structure of the general \name
    library. (a) The proportion of declarations in each
    top-level category. (b) The reusable code
    \texttt{List.isPalindrome} and four representative theorems in its
    declaration family.}
    \label{fig:veribase-library-composition}
\end{figure}

The general library contains 10 top-level categories, each further divided into
topic-specific modules, for a total of 74 modules. For example, the
\texttt{List} category includes \texttt{Sort}, \texttt{Count}, and
\texttt{Range}, while the \texttt{String} category includes
\texttt{Substrings}, \texttt{NumericConversion}, and
\texttt{Normalization}. The library contains about 54K lines of Lean~4 code.
Figure~\ref{fig:veribase-library-composition}(a) shows that the distribution is
highly skewed: \texttt{List} accounts for 53.6\% of all declarations, followed
by \texttt{String} with 13.6\% and \texttt{Nat} with 11.5\%. This distribution
reflects the prevalence of list processing in software-verification tasks,
while the remaining categories provide reusable support for trees, arrays,
characters, finite collections, grids, integers, and others.

The library has a code-centered structure, comprising 918
theorem families in which a code declaration is accompanied by one or more theorems
that directly reference it. These theorem families capture
reusable properties such as membership preservation, output length,
permutation, ordering, and bounds.
Figure~\ref{fig:veribase-library-composition}(b) illustrates this structure
through \texttt{List.isPalindrome}. The code is accompanied
by theorems that characterize the code's  Boolean result, establish a base case,
construct new palindromes, and show invariance under reversal. These interfaces
allow subsequent proofs to reason about \texttt{isPalindrome} without
repeatedly unfolding its implementation.

\subsection{Benchmark-Specific Libraries}

\name constructs four benchmark-specific libraries from the construction
subsets of Verina~\cite{ye2025verina}, FVAPPS~\cite{dougherty2025proving},
VeriContest~\cite{xie2026vericontest}, and VeriSoftBench~\cite{xin2026verisoftbench}.
Table~\ref{tab:benchmark-specific-libraries} summarizes their sizes and
representative contents. These libraries provide code and theorems tailored
to the operations and properties encountered in their source benchmarks.

\begin{table}[t]
    \centering
    \caption{Contents of the four benchmark-specific libraries, including the
    numbers of definitions and theorems and representative content.}
    \label{tab:benchmark-specific-libraries}
    \vspace{-2mm}
    \TableStyle
    \begin{tabular}{@{}lrrp{0.43\linewidth}@{}}
        \toprule
        \textbf{Library} & \textbf{Definitions} & \textbf{Theorems} & \textbf{Representative content} \\
        \midrule
        Verina & 49 & 114 & Sorting, aggregation, extremum selection \\
        FVAPPS & 188 & 451 & Counting, list windows, string processing \\
        VeriContest & 433 & 659 & Prefix sums, contiguous segments, arithmetic \\
        VeriSoftBench & 0 & 811 & Evaluation semantics, program execution \\
        \bottomrule
    \end{tabular}
\end{table}

Like the general library, the Verina, FVAPPS, and VeriContest libraries are
organized into top-level categories by data type, each further divided into
topic-specific modules stored in separate files. These libraries also have a
code-centered structure, with related theorems organized around code.
The VeriSoftBench library is organized into top-level categories by referenced repositories, each further divided into topic-specific modules stored in
separate files. VeriSoftBench requires only proof generation.
Its generated library provides 811 theorems but no executable code. The code to be verified
is fixed by the referenced repositories, so these tasks benefit from
additional theorems about that code rather than alternative code implementations.
These theorems are still organized around code in the referenced libraries,
so the VeriSoftBench library also has a code-centered structure.
The four libraries can be used
independently or combined with the general library to provide additional
code and theorems relevant to a particular benchmark.

\subsection{Construction Cost and Ecosystem Position}
\noindent\textbf{Construction Cost.}
The estimated API cost of constructing the general library is approximately \$431;
the corresponding costs for the Verina, FVAPPS, VeriContest, and VeriSoftBench
libraries are approximately \$11, \$103, \$87, and \$19, respectively.
These estimates cover all three construction phases, including failed attempts
and retries. Excluding the proving phase, as when successful solutions are already
available, the estimated construction costs are approximately \$62 for the general
library, and \$2, \$15, \$12, and \$3 for
the Verina, FVAPPS, VeriContest, and VeriSoftBench libraries, respectively.

\noindent\textbf{Position in the Verification-Library Ecosystem.}
Across verification ecosystems, reusable libraries %
provide software components
with the specifications, lemmas, theorems, and other artifacts that support %
reasoning about the software components. 
Building such shared infrastructure, however,
requires substantial expert effort. Library developers must identify recurring
abstractions, design stable interfaces, prove the associated properties,
reconcile overlapping components, and organize the results into maintainable
modules. Even when individual projects or tasks already contain many successful
verifications, turning these verification artifacts into a coherent library remains costly, especially in a new domain.
\name automates library construction for verification within Lean. The generated code can
be reused directly in Lean programs, while the associated theorems support proofs about
these implementations and repository-specific formalizations. For software
written in other languages, such as C and  Rust, reuse additionally requires
establishing a semantic connection between the source programs and their Lean
representations.
This automation can also complement manual curation: \name can bootstrap an
initial library whose interfaces, organization, and coverage are subsequently
refined by experts. This bootstrapping provides an automated starting point for building and
extending verification libraries while retaining a role for expert review and
maintenance.

%% file: tex/005_experiments.tex
\section{Evaluation}
We conduct extensive experiments to evaluate the effectiveness of the \name libraries (including the general library and benchmark-specific libraries).
Specifically, we focus on the following research questions (RQs):

\begin{itemize}[leftmargin=*]
    \item \textbf{RQ1:} How effectively do the  \name libraries  improve task success rates?
    \item \textbf{RQ2:} How do the  \name libraries affect task completion time and API cost?
    \item \textbf{RQ3:} How does general-library size affect task success rates? %
    \item \textbf{RQ4:} How do the  \name libraries  compare with retrieving 
    raw successful solutions directly, and how do code, theorems,
    generalization, and harvesting contribute to their effectiveness?
\end{itemize}

\subsection{Experimental Setup}
\label{sec:experimental-setup}

\noindent\textbf{Datasets.}
We consider four
software verification benchmarks. The first three contain single-file tasks
requiring both code generation and proof generation, whereas the last contains
repository-level tasks requiring proof generation only.
\begin{itemize}[leftmargin=*]
    \item \textbf{Verina:} Verina~\cite{ye2025verina} contains 189 manually
    curated Lean coding tasks at basic and advanced difficulty levels. To make the test split more challenging, we use the
    108 basic tasks to construct the benchmark-specific library and the 81 advanced tasks as the test set.
    \item \textbf{FVAPPS:} FVAPPS~\cite{dougherty2025proving} contains 4,715
    Lean formalizations of general-purpose coding-interview problems from APPS~\cite{Hendrycks2021MeasuringCC}. We randomly select 100 problems as the test set and select 1,000
    other tasks to construct the benchmark-specific library.
    When selecting test tasks, we exclude those sharing %
    an identifier with any of the 4,191 tasks in the general library's construction
    corpus.
    \item \textbf{VeriContest:} VeriContest~\cite{xie2026vericontest} contains
    946 competitive-programming problems. We randomly select 100 problems as the test set and use the remaining 846 tasks to construct the benchmark-specific library.
    When selecting test tasks, we exclude those sharing an original LeetCode
    problem identifier with any of the 4,191 tasks in the general library's
    construction corpus.
    \item \textbf{VeriSoftBench:} VeriSoftBench~\cite{xin2026verisoftbench}
    contains 500 repository-level Lean~4 proof obligations spanning compiler
    correctness, programming-language semantics and type systems, verification
    frameworks, and applied verification.
    After aligning the Lean versions, we retain 283 tasks. We
    randomly select 183 tasks to construct the benchmark-specific library and use the remaining 100 as the test set.
\end{itemize}

\noindent\textbf{Metrics.}
We use the following three metrics. 
\begin{itemize}[leftmargin=*]
    \item \textbf{Task success rate:} The fraction of tasks for which the agent
    produces all artifacts required by the benchmark within the inference
    budget.
    \item \textbf{Average completion time:} The average wall-clock time spent
    by the agent on each task, with timed-out attempts counted as 1,800 seconds.
    \item \textbf{Average API cost:} The average inference cost in USD over all
    tasks, computed from recorded uncached-input, cache-hit, cache-creation, and
    output token counts using public API prices. 
\end{itemize}

\noindent\textbf{Settings.}
We use Claude Code as the agent harness, which manages tool use and interaction with the verification environment, with three backend LLMs: DeepSeek V4 Pro~\cite{xu2026deepseek},
GLM 5.2~\cite{glm5team2026glm5vibecodingagentic}, and Qwen 3.7 Max~\cite{qwen37}. For each test task, the agent follows the code-first,
proof-second workflow of the proving Phase, iteratively revising the generated code and
proof based on feedback from Lean. Each task has a time budget of 1,800 seconds.
The general library is shared across benchmarks, whereas each benchmark-specific
library is used for its corresponding benchmark. 
In settings with library access, we provide a
simple search program that retrieves relevant declarations by combining semantic
similarity with keyword matching for agent-generated queries. All libraries
remain fixed during evaluation; no declarations are added, modified, or removed. %

\smallskip
\noindent\textbf{Validation Protocol.}
A task is considered successful only if its final artifact preserves the original task components used for evaluation. 
These components include the code specifications
(preconditions, postconditions, and their supporting definitions) and the statement of the target theorem. The final artifact must also compile in Lean, and the proof and its transitive dependencies must contain no unproved placeholders
or unauthorized axioms. 

\begin{figure}[t]
    \centering
    \includegraphics[width=0.9\linewidth]{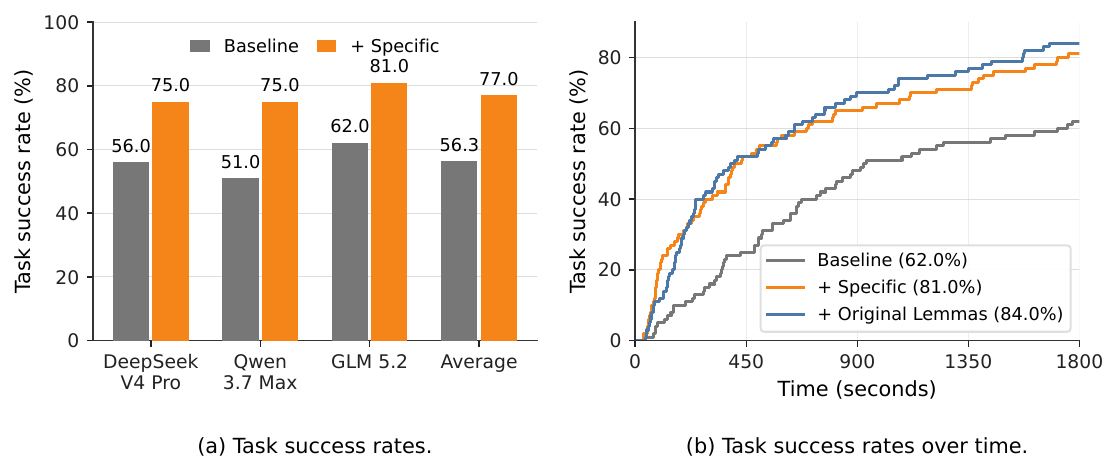}
    \vspace{-2mm}
    \caption{Repository knowledge recovery on VeriSoftBench. \emph{Baseline}
    uses the environment with theorems removed; \emph{+ Specific} adds the
    \name-constructed library; \emph{+ Original Lemmas} restores the original
    lemmas instead. (a) Task success rates and their macro-average across three
    backends. (b) Task success rates over time for GLM 5.2.}
    \label{fig:repository-knowledge-recovery}
    \Description{The bar chart compares baseline and specific-library success
    rates for three backends and their average. The step chart compares three
    GLM 5.2 settings over time, ending at 62, 81, and 84 percent.}
\end{figure}

\subsection{RQ1: Effectiveness of \name Libraries}

\begin{table}[t]
    \centering
    \caption{Task success rates (\%) across three benchmarks and three backend LLMs.  \emph{Baseline} means the standalone proving, \emph{+Gen.} means when the general library is available, and
    \emph{+Gen.\& Spec.} means both the general library and the benchmark-specific library are available. Absolute changes compared to \emph{Baseline} are shown in the brackets.} 
    \label{tab:library-effectiveness}
    \vspace{-2mm}
    \TableStyle
    \renewcommand{\arraystretch}{1.0}
    \begin{tabular}{@{}ll
        S[table-format=2.1]
        S[table-format=2.1]@{\hspace{0.25em}}l
        S[table-format=2.1]@{\hspace{0.25em}}l@{}}
        \toprule
        \textbf{Benchmark}
        & \textbf{Backend LLM}
        & \multicolumn{1}{c}{\textbf{Baseline}}
        & \multicolumn{2}{c}{\textbf{+Gen.}}
        & \multicolumn{2}{c}{\textbf{+Gen. \& Spec.}} \\
        \midrule
        Verina
        & DeepSeek V4 Pro & 66.7 & 80.2 & \goodchange{+13.6} & 77.8 & \goodchange{+11.1} \\
        & GLM 5.2 & 58.0 & 74.1 & \goodchange{+16.0} & 88.9 & \goodchange{+30.9} \\
        & Qwen 3.7 Max & 64.2 & 74.1 & \goodchange{+9.9} & 71.6 & \goodchange{+7.4} \\
        \midrule
        FVAPPS
        & DeepSeek V4 Pro & 54.0 & 71.0 & \goodchange{+17.0} & 74.0 & \goodchange{+20.0} \\
        & GLM 5.2 & 59.0 & 66.0 & \goodchange{+7.0} & 76.0 & \goodchange{+17.0} \\
        & Qwen 3.7 Max & 57.0 & 63.0 & \goodchange{+6.0} & 66.0 & \goodchange{+9.0} \\
        \midrule
        VeriContest
        & DeepSeek V4 Pro & 70.0 & 76.0 & \goodchange{+6.0} & 85.0 & \goodchange{+15.0} \\
        & GLM 5.2 & 69.0 & 78.0 & \goodchange{+9.0} & 85.0 & \goodchange{+16.0} \\
        & Qwen 3.7 Max & 69.0 & 75.0 & \goodchange{+6.0} & 82.0 & \goodchange{+13.0} \\
        \bottomrule
    \end{tabular}
\end{table}

We compare three settings. \emph{Baseline} uses no %
library;
\emph{+Gen.} allows the agent to retrieve declarations from the general
library; and \emph{+Gen. \& Spec.} allows the agent to retrieve
declarations from both the general and benchmark-specific libraries.   

\noindent\textbf{Effectiveness of the General Library.}
As shown in Table~\ref{tab:library-effectiveness}, the general library improves
task success rates in all nine benchmark--backend combinations. Overall, the
macro-average increases from 63.0\% to 73.0\%.
These results show that the general library provides reusable code and
theorems that transfer across task distributions and backend LLMs.

\noindent\textbf{Extra Benefits of Benchmark-specific Library.}
As shown in Table~\ref{tab:library-effectiveness}, composing the general and
benchmark-specific libraries raises the overall macro-average from 73.0\% to
78.5\%. This combined setting improves seven of the nine benchmark--backend combinations; the remaining
two combinations show only minor regressions of 2.4 and 2.5 points.
GLM 5.2 shows particularly large gains. %
These results show that benchmark-specific libraries can complement the
general library with reusable code and theorems from the target domain,
further improving verification performance.

\noindent\textbf{Effectiveness on Repository-level Tasks.}
VeriSoftBench references software repositories with extensive human-developed
supporting theorems, which serve a role analogous to our constructed libraries.
We assess whether an automatically constructed library can provide comparable
verification support. We remove human-developed supporting theorems,
primarily those tagged by VeriSoftBench as \texttt{repo\_lemmas} and
\texttt{used\_local\_lemmas}. We then use \name to construct a library from the 183 construction tasks in this modified environment.
\emph{Baseline} uses the modified environment alone; \emph{+ Specific} adds
this constructed library; and \emph{+ Original Lemmas} uses the original
environment without adding the constructed library. The general library is not used.
Figure~\ref{fig:repository-knowledge-recovery}(a) shows that \emph{+ Specific}
improves success rates across all three backends, raising the average from
56.3\% to 77.0\%. For GLM 5.2, Figure~\ref{fig:repository-knowledge-recovery}(b) shows that
\emph{+ Specific} raises the success rate from 62.0\% to 81.0\%, close to
84.0\% under \emph{+ Original Lemmas}, suggesting comparable quality
between automatic and human-developed libraries for repository-level verification.

\begin{rqsummary}
The general \name library consistently improves task success rates, with a
macro-average absolute gain of 10.0\% over baseline. Adding benchmark-specific libraries increases the %
gain to 15.5\%. %
On VeriSoftBench, the constructed library provides verification support close
to that of the original human-developed theorems, suggesting comparable library quality.
\end{rqsummary}

\subsection{RQ2: Inference Efficiency of \name Libraries}
\begin{table}[t]
    \centering
    \caption{Average completion time (seconds) and API cost
    (USD) across three benchmarks and three
    backend LLMs. Performance changes compared to \emph{Baseline} are shown in the brackets.}
    \label{tab:general-library-efficiency}
    \vspace{-2mm}
    \TableStyle
    \footnotesize
    \renewcommand{\arraystretch}{1.0}
    \setlength{\tabcolsep}{3pt}
    \begin{tabular}{@{}ll
        S[table-format=4.1]
        S[table-format=4.1]@{\hspace{0.25em}}l
        S[table-format=4.1]@{\hspace{0.25em}}l
        S[table-format=1.3]
        S[table-format=1.3]@{\hspace{0.25em}}l
        S[table-format=1.3]@{\hspace{0.25em}}l@{}}
        \toprule
        \textbf{Benchmark}
        & \textbf{Backend LLM}
        & \multicolumn{5}{c}{\textbf{Time (s)}}
        & \multicolumn{5}{c}{\textbf{Cost (\$)}} \\
        \cmidrule(lr){3-7}\cmidrule(l){8-12}
        & & \multicolumn{1}{c}{\textbf{Baseline}} & \multicolumn{2}{c}{\textbf{+Gen.}}
        & \multicolumn{2}{c}{\textbf{+Gen. \& Spec.}}
        & \multicolumn{1}{c}{\textbf{Baseline}} & \multicolumn{2}{c}{\textbf{+Gen.}}
        & \multicolumn{2}{c}{\textbf{+Gen. \& Spec.}} \\
        \midrule
        Verina
        & DeepSeek V4 Pro    & 1123.7 & 883.5 & \goodchange{-21.4\%} & 771.2 & \goodchange{-31.4\%} & 0.071 & 0.084 & \badchange{+18.3\%} & 0.102 & \badchange{+43.7\%} \\
        & GLM 5.2            & 1274.0 & 1020.3 & \goodchange{-19.9\%} & 597.5 & \goodchange{-53.1\%} & 1.204 & 1.084 & \goodchange{-10.0\%} & 0.807 & \goodchange{-33.0\%} \\
        & Qwen 3.7 Max       & 1213.6 & 1005.7 & \goodchange{-17.1\%} & 978.3 & \goodchange{-19.4\%} & 1.153 & 0.927 & \goodchange{-19.6\%} & 0.955 & \goodchange{-17.2\%} \\
        \midrule
        FVAPPS
        & DeepSeek V4 Pro    & 1150.1 & 1008.1 & \goodchange{-12.3\%} & 895.6 & \goodchange{-22.1\%} & 0.072 & 0.096 & \badchange{+33.3\%} & 0.114 & \badchange{+58.3\%} \\
        & GLM 5.2            & 1229.0 & 1109.2 & \goodchange{-9.7\%} & 837.0 & \goodchange{-31.9\%} & 1.089 & 1.211 & \badchange{+11.2\%} & 0.888 & \goodchange{-18.5\%} \\
        & Qwen 3.7 Max       & 1223.7 & 1126.3 & \goodchange{-8.0\%} & 1103.9 & \goodchange{-9.8\%} & 1.186 & 1.057 & \goodchange{-10.9\%} & 1.287 & \badchange{+8.5\%} \\
        \midrule
        VeriContest
        & DeepSeek V4 Pro    & 893.6 & 815.9 & \goodchange{-8.7\%} & 606.0 & \goodchange{-32.2\%} & 0.060 & 0.073 & \badchange{+21.7\%} & 0.076 & \badchange{+26.7\%} \\
        & GLM 5.2            & 1012.7 & 919.7 & \goodchange{-9.2\%} & 680.7 & \goodchange{-32.8\%} & 0.872 & 0.864 & \goodchange{-0.9\%} & 0.751 & \goodchange{-13.9\%} \\
        & Qwen 3.7 Max       & 931.6 & 837.7 & \goodchange{-10.1\%} & 777.9 & \goodchange{-16.5\%} & 0.723 & 0.775 & \badchange{+7.2\%} & 0.929 & \badchange{+28.5\%} \\
        \bottomrule
    \end{tabular}
\end{table}

We evaluate how the general library and  benchmark-specific libraries
affect average completion time and API cost on the three single-file benchmarks.

\noindent\textbf{Completion Time.}
We average over all test tasks, counting timeouts as 1,800 seconds.
As shown in Table~\ref{tab:general-library-efficiency}, the general library
reduces average completion time in all nine benchmark--backend combinations,
from 1,116.9 to 969.6 seconds on average, a 13.2\% reduction. Composing the
general library with benchmark-specific libraries further reduces completion
time in all nine combinations, to 805.3 seconds on average, an overall 27.9\%
reduction relative to the baseline. These consistent reductions across all nine
combinations show that library access improves verification efficiency alongside
task success rates. Benchmark-specific libraries further accelerate verification
beyond the gains provided by the general library.

\noindent\textbf{API Cost.}
In Table~\ref{tab:general-library-efficiency}, the API cost
varies across backend LLMs. Under \emph{+Gen.}, costs increase by
18.3--33.3\% for DeepSeek V4 Pro, and show mixed changes for GLM 5.2 and Qwen
3.7 Max. 
Under \emph{+Gen. \& Spec.}, costs decrease by 13.9--33.0\% for
GLM 5.2, increase by 26.7--58.3\% for DeepSeek V4 Pro, and remain mixed for
Qwen 3.7 Max.
Library retrieval adds input context, whereas reusing
library declarations can reduce subsequent generation and revision costs.
GLM 5.2 shows reductions in both input and output token
consumption. For Qwen and DeepSeek, lower output consumption is sometimes accompanied by increased input consumption, which can offset the savings. This trade-off, together with model-specific token prices, helps explain the differing cost trends.

\begin{rqsummary}
\name libraries consistently and substantially reduce completion time, with
benchmark-specific libraries  providing further reductions beyond the general
library. However, these efficiency gains do not always translate into lower
API cost, wich varies across backend LLMs.
\end{rqsummary}

\subsection{RQ3: Impact of General-Library Size}
We evaluate how general-library size influences task performance by constructing general libraries from nested subsets of 100, 500, and all
4,191 construction tasks. We evaluate these libraries, together with a no-library baseline, on Verina using DeepSeek V4 Pro under the same inference budget.

\begin{figure}[t]
    \centering
    \begin{minipage}[t]{0.46\linewidth}
        \centering
        \includegraphics[width=\linewidth]{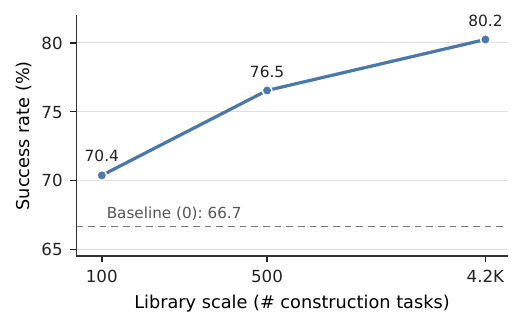}
        \small (a) Final task success rate.
    \end{minipage}
    \hspace{0.02\linewidth}
    \begin{minipage}[t]{0.46\linewidth}
        \centering
        \includegraphics[width=\linewidth]{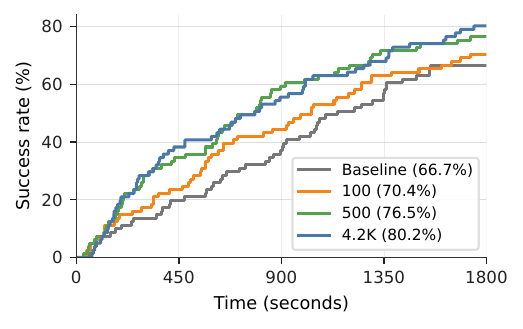}
        \small (b) Task success rate over time.
    \end{minipage}
    \caption{Impact of construction-corpus size on the effectiveness of the
    general library, evaluated on Verina using DeepSeek V4 Pro. (a) Final
    task success rates for the no-library baseline and libraries built from
    100, 500, and 4,191 library-construction tasks. (b) Task success rates over time
    under the same settings.}
    \label{fig:library-scale}
\end{figure}

Figure~\ref{fig:library-scale}(a) shows task success rates, %
indicating extra but diminishing returns as the corpus grows.
In Figure~\ref{fig:library-scale}(b), the 500-task and full libraries follow
similar trajectories, outperforming the 100-task library and then the baseline.
Efficiency thus largely saturates beyond 500 construction tasks, while final
task coverage continues to improve.

A larger corpus exposes \name to more reusable code and theorems, increasing
the chance of finding declarations that match a task. Reuse reduces generation
from scratch and subsequent repair, improving both task success rates and
efficiency.
Diminishing returns reflect the limited range of common data types, such as
lists, strings, and natural numbers, and operations within each type, such as
counting, folding, and sorting. Once these type--operation families are covered,
further harvesting tends to yield narrower code and theorems with fewer reuse
opportunities. This coverage pattern suggests a long-tail distribution: common families are
harvested early, while later additions serve more specialized tasks.

\begin{rqsummary}
Increasing the general-library size improves the task success rate, but with diminishing
returns. Inference efficiency improves substantially at smaller scales and begins
to saturate as the number of library-construction tasks grows beyond 500.
\end{rqsummary}

\subsection{RQ4: Ablation Study}
We compare the combined general and VeriContest-specific library with five
alternative settings on VeriContest using GLM 5.2.
\emph{Raw Solution Retrieval} retrieves original successful solutions from both
construction corpora. \emph{w/o Code} removes code and associated theorems;
\emph{w/o Theorems} removes theorems while preserving code.
\emph{w/o Generalization} disables generalization both before and during
harvesting. \emph{w/o Harvesting} mechanically assembles the frequency-filtered
generalized candidates and their dependencies into a compilable library.
Only the processing and presentation of construction material differ; test tasks,
backend LLM, retrieval scoring, and per-task inference budget remain fixed across settings.

\begin{table}[t]
    \centering
    \caption{Library composition and task success rates (\%) of different
    variants on VeriContest using GLM 5.2. Counts for \emph{Raw Solution Retrieval} denote
    indexed source declarations rather than a compiled library.}
    \label{tab:ablation-effectiveness}
    \TableStyle
    \footnotesize
    \renewcommand{\arraystretch}{1.0}
    \begin{tabular}{@{}lrrr@{}}
        \toprule
        \textbf{Configuration}
        & \textbf{Definitions}
        & \textbf{Theorems}
        & \shortstack{\textbf{Task success rate} \textbf{(\%)}} \\
        \midrule
        Baseline           & --     & --    & 69.0 \\
        Raw Solution Retrieval & 24,790 & 12,813 & 72.0 \\
        w/o Code           & 0      & 1,399 & 74.0 \\
        w/o Theorems       & 2,042  & 0     & 75.0 \\
        w/o Generalization & 7,006  & 2,958 & 71.0 \\
        w/o Harvesting     & 10,056 & 7,089 & 65.0 \\
        \textbf{Full \name} & \textbf{2,042} & \textbf{3,898} &
        \textbf{85.0} \\
        \bottomrule
    \end{tabular}
\end{table}

\noindent\textbf{Comparison with Raw Solution Retrieval.}
Retrieving raw successful solutions achieves a task success rate of 72.0\%,
compared with 85.0\% for the full library. Raw solutions retain task-specific
interfaces and assumptions, and thus require more adaptation, showing that
direct access to original code and proofs does not match a generalized and
organized library.

\noindent\textbf{Effect of Joint Code-Theorem Harvesting.}
In Table~\ref{tab:ablation-effectiveness}, \emph{w/o Code} and \emph{w/o Theorems}
achieve 74.0\% and 75.0\%, respectively, versus 69.0\% for Baseline and 85.0\%
for Full. Code alone leaves properties to be proved; standalone theorems omit
code-dependent knowledge. Both gaps support retaining code and associated theorems together.

\noindent\textbf{Effect of Generalization.}
Removing generalization reduces the task success rate from 85.0\% to
71.0\%, despite retaining 9,964 declarations versus 5,940 in the full library.
This result shows that generalization improves applicability and that library
effectiveness depends on more than the number of declarations in the library.

\noindent\textbf{Effect of Harvesting.}
Mechanical aggregation causes the largest reduction, from 85.0\% to 65.0\%,
despite retaining 17,145 declarations versus 5,940 in the full library.
Harvesting filters covered or low-value candidates, merges overlapping
interfaces, and organizes the library, producing a smaller library with less
retrieval noise.

\begin{rqsummary}
Direct retrieval of raw successful solutions is less effective than using
the constructed library. Retaining code together with its associated
theorems is more effective than retaining either alone. Generalization broadens
the applicability of the code and its associated theorems, and harvesting curates and consolidates the library;
removing either reduces task success rates.
\end{rqsummary}

\input{tex/006_analysis}

%% file: tex/006_analysis.tex
\subsection{Library Reuse Analysis}
\label{sec:library-reuse-analysis}

Across three single-file benchmarks and three LLM backends, we analyze 661
successful task--backend pairs under \emph{+Gen.\ \& Spec.}: 117 explicitly
reference harvested code, 173 reference harvested theorems, and 83 reference
code together with its associated theorems, providing direct evidence of joint
reuse. We count only fully qualified references in generated implementations
and proofs, excluding comments, imports, given specifications, and local
declarations. These counts are conservative lower bounds because they omit
unqualified references, dot notation, and implicit uses of library declarations by tactics.

\noindent\textbf{Case Study of Cross-task Reuse.}
The construction task \texttt{LeetCode\_3168} computes the maximum occupancy of
a waiting room. Through generalization and Harvesting, its solution contributes to %
the \texttt{List.listMax} declaration family, including the two theorems shown in
Figure~\ref{fig:cross-task-reuse}. 
Verina's \texttt{advanced\_38} instead
maximizes interval coverage after removing one interval; its generated solution
reuses \texttt{List.listMax}  and both associated theorems to prove attainment and optimality. GLM 5.2
times out without library access but completes the task in 213 seconds with access to the combined general and Verina-specific library.

\begin{figure}[t]
    \centering
    \begin{CodePair}
    \begin{CodePairLeft}
    \begin{LeanCode}
-- Harvested library interface
def List.listMax (xs : List Nat) : Nat := ...

lemma List.listMax_map_range_ge
    (f : Nat -> Nat) (n j : Nat) (hj : j < n) :
    f j <= List.listMax ((List.range n).map f) := ...

lemma List.listMax_map_range_exists_eq
    (f : Nat -> Nat) (n : Nat) (hn : 1 <= n) :
    exists i < n,
      f i = List.listMax ((List.range n).map f) := ...
    \end{LeanCode}
    \end{CodePairLeft}
    \begin{CodePairRight}
    \begin{LeanCode}
-- Implementation in Verina advanced_38
List.listMax ((List.range intervals.length).map
  (coverageAfterRemoval intervals))

-- Proof excerpt
have hex := List.listMax_map_range_exists_eq
  (coverageAfterRemoval intervals)
  intervals.length hlen
have hge := List.listMax_map_range_ge
  (coverageAfterRemoval intervals)
    \end{LeanCode}
    \end{CodePairRight}
    \end{CodePair}
    \caption{Code and theorems derived from a chair-allocation task (left) are
    reused for interval-coverage verification (right).}
    \label{fig:cross-task-reuse}
    \Description{A generic maximum-over-range function and two theorems derived
    from a chair-allocation task are used to implement and prove an interval
    coverage optimization task.}
\end{figure}

%% file: tex/006_threats.tex
\section{Threats to Validity}
\label{sec:threats}
\noindent\textbf{Overlap between Construction and Test Tasks.}
Task overlap could inflate the measured library benefits. We separate test
tasks from both construction corpora and, for FVAPPS and VeriContest, exclude
test tasks sharing source identifiers with the general construction corpus.
These controls reduce overlap risk, so we expect its impact on the reported
gains to be limited.

\noindent\textbf{Variability in LLM Performance.}
Generation randomness and service load can affect success rates and efficiency
independently of library use. We evaluate three backends and, for each model,
use the same inference budget and concurrency across settings while avoiding
rate-limit errors. These measures reduce model dependence and variation in
execution conditions. Although fluctuations remain, we expect their impact on
our overall conclusions to be limited.

\noindent\textbf{Dependence on the Retrieval Setup.}
Retrieval quality affects whether agents find relevant declarations. We fix
the scoring technique and number of results returned per library across settings;
additional checks show that retrieval can effectively identify relevant
declarations. Although this setup may not make optimal use of the libraries,
its consistency across settings leads us to expect only a small impact on
their comparison.

%% file: tex/007_related_work.tex
\section{Related Work}
\subsection{Library Learning}
Library learning has been studied extensively in program synthesis, where
reusable abstractions can both simplify existing programs and accelerate the
synthesis of new ones.
DreamCoder~\cite{ellis2021dreamcoder} iteratively builds a library of reusable
program components by solving tasks, refactoring successful programs, and
training a neural search policy in a wake--sleep loop.
\textsc{Stitch}~\cite{bowers2023top} uses corpus-guided top-down synthesis
to identify shared abstractions efficiently. LILO~\cite{grand2024lilo}
combines LLM-guided synthesis, \textsc{Stitch}-based compression, and automatic
documentation to construct interpretable, domain-specific libraries.
These approaches learn reusable code abstractions, whereas software verification additionally requires reusable theorems that specify and reason about the behavior of reusable code components.

Library learning has also been explored in formal theorem proving.
LEGO-Prover~\cite{wang2024lego} maintains a growing library of verified
lemmas as reusable skills that can be retrieved, created, and evolved during
proof construction. However, a subsequent
study~\cite{berlot2025llm} reports limited lemma reuse
and no measurable performance improvement under controlled computational
budgets.
Seed-Prover~\cite{chen2025seed} adopts lemma-style proving and maintains
a lemma pool for each difficult problem, allowing verified lemmas from
different inference runs to be retrieved, refined, and combined.
DreamProver~\cite{zhang2026dreamprover} uses a wake--sleep process to prove
training theorems and abstract, refine, and consolidate candidate lemmas into
a transferable library.
Together, these lines of work establish library learning for reusable program
abstractions and proof lemmas. In contrast, \name focuses on the coordinated transformation
of executable code and its associated theorems into software-verification
libraries. It generalizes implementation interfaces, adapts theorem statements
and proofs to those interfaces, and consolidates overlapping declaration families
across tasks while updating their dependencies. The resulting libraries support
reuse in both implementation generation and correctness proofs.

\subsection{Reusable Libraries for Software Verification}
Formal verification ecosystems provide reusable code, specifications,  and
theorems. Why3~\cite{filliatre2013why3} provides logical theories and basic
program data structures, while Dafny~\cite{dafnystandardlibraries} offers
pre-verified standard libraries for common programming operations. Verus's
\texttt{vstd}~\cite{lattuada2023verus} combines proof utilities with specified runtime
functionality, and the Isabelle Collections Framework~\cite{lammich2010isabelle}
provides verified collection data structures through data refinement.
Rocq's \texttt{std++}~\cite{stdpp} extends its standard library with definitions and
theorems for common data structures, including lists, finite maps, finite sets,
and finite multisets. VSTlib~\cite{appel2024vstlib} specifies standard C
components such as memory allocation, atomics, locks, and threads; some
implementations are verified, while others are axiomatized system-library
interfaces.
Together, these libraries allow users to invoke  standard library components under
formal specifications rather than reimplement and reverify them.

In Lean~4, \texttt{Std}, \texttt{Batteries}, and \texttt{Mathlib}
~\cite{moura2021lean,mathlib2020} provide foundational
data structures and
supporting theorems. However, these libraries primarily provide basic,
general-purpose infrastructure and are not sufficiently tailored to diverse
software-verification tasks, which require domain-specific implementations
and surrounding theorem families. CSLib~\cite{barrett2026cslib} aims to
address this gap with a Mathlib-like repository of Computer Science concepts,
algorithms, data structures, and verified properties. It also targets
Lean-based reasoning about everyday imperative code and the composition of
reliable systems from pre-verified components. However, CSLib remains at an early stage of building this shared infrastructure.

These previous efforts demonstrate the value of reusable verification libraries, but
their contents are primarily designed and maintained manually by domain
experts. In contrast, \name automates part of this library-development process by extracting
code and associated theorems from completed verification tasks, generalizing
their interfaces together, and integrating overlapping families into a compiled
library. These libraries can serve as starting points for further expert curation.

%% file: tex/008_conclusion.tex
\section{Conclusion}
In this paper, we have presented \name, a framework for automatically constructing reusable
software-verification libraries from successful task solutions. Its three-phase
pipeline proves seed tasks, generalizes implementations together with their
associated theorems, and consolidates overlapping declaration families across
tasks. \name updates dependent declarations as interface change and checks the
resulting code and proofs in Lean.
Using \name, we have constructed a general theorem
library for Lean~4 software verification, as well as benchmark-specific libraries. 
Experiments on Verina, FVAPPS, VeriContest, and VeriSoftBench demonstrate that
the resulting libraries improve task success rates and consistently reduce
completion time, while their effect on API cost
varies across backend LLMs. 
These results support coordinated generalization and integration of code and
theorems as an approach to turning task-level verification effort into reusable
library interfaces for subsequent tasks.